\documentclass[12pt]{article}
\input psfig.sty
\input axodraw.sty

\def\Re{{\cal R \mskip-4mu \lower.1ex \hbox{\it e}\,}}
\def\Im{{\cal I \mskip-5mu \lower.1ex \hbox{\it m}\,}}

\def\sub#1{_{\lower.25ex\hbox{$\scriptstyle#1$}}}
\def\tev{\,{\ifmmode\mathrm {TeV}\else TeV\fi}}
\def\gev{\,{\ifmmode\mathrm {GeV}\else GeV\fi}}
\def\mev{\,{\ifmmode\mathrm {MeV}\else MeV\fi}}
\def\mpl{\ifmmode \overline M_{Pl}\else $\overline M_{Pl}$\fi}
\def\to{\rightarrow}

\def\subw{_{\rm w}}
\def\mh{\ifmmode m\sbl H \else $m\sbl H$\fi}
\def\mch{\ifmmode m_{H^\pm} \else $m_{H^\pm}$\fi}
\def\mt{\ifmmode m_t\else $m_t$\fi}
\def\mc{\ifmmode m_c\else $m_c$\fi}
\def\mz{\ifmmode M_Z\else $M_Z$\fi}
\def\mw{\ifmmode M_W\else $M_W$\fi}
\def\mws{\ifmmode M_W^2 \else $M_W^2$\fi}
\def\mhs{\ifmmode m_H^2 \else $m_H^2$\fi}   
\def\mzs{\ifmmode M_Z^2 \else $M_Z^2$\fi}
\def\mts{\ifmmode m_t^2 \else $m_t^2$\fi}
\def\mcs{\ifmmode m_c^2 \else $m_c^2$\fi}
\def\mchs{\ifmmode m_{H^\pm}^2 \else $m_{H^\pm}^2$\fi}
\def\ztwo{\ifmmode Z_2\else $Z_2$\fi}
\def\zone{\ifmmode Z_1\else $Z_1$\fi}
\def\mtwo{\ifmmode M_2\else $M_2$\fi}
\def\mone{\ifmmode M_1\else $M_1$\fi}
\def\tb{\ifmmode \tan\beta \else $\tan\beta$\fi}
\def\xw{\ifmmode x\subw\else $x\subw$\fi}
\def\ch{\ifmmode H^\pm \else $H^\pm$\fi}
\def\lum{\ifmmode {\cal L}\else ${\cal L}$\fi}
\def\inpb{\,{\ifmmode {\mathrm {pb}}^{-1}\else ${\mathrm {pb}}^{-1}$\fi}}
\def\infb{\,{\ifmmode {\mathrm {fb}}^{-1}\else ${\mathrm {fb}}^{-1}$\fi}}
\def\epem{\ifmmode e^+e^-\else $e^+e^-$\fi}
\def\ppb{\ifmmode \bar pp\else $\bar pp$\fi}
\def\bsg{\ifmmode B\to X_s\gamma\else $B\to X_s\gamma$\fi}
\def\bsll{\ifmmode B\to X_s\ell^+\ell^-\else $B\to X_s\ell^+\ell^-$\fi}
\def\bstt{\ifmmode B\to X_s\tau^+\tau^-\else $B\to X_s\tau^+\tau^-$\fi}
\def\lamt{\ifmmode \tilde\lambda\else $\tilde\lambda$\fi}
\def\shat{\ifmmode \hat s\else $\hat s$\fi}
\def\that{\ifmmode \hat t\else $\hat t$\fi}
\def\uhat{\ifmmode \hat u\else $\hat u$\fi}

\newskip\zatskip \zatskip=0pt plus0pt minus0pt
\def\matth{\mathsurround=0pt}
\def\lsim{\mathrel{\mathpalette\atversim<}}
\def\gsim{\mathrel{\mathpalette\atversim>}}
\def\atversim#1#2{\lower0.7ex\vbox{\baselineskip\zatskip\lineskip\zatskip
  \lineskiplimit 0pt\ialign{$\matth#1\hfil##\hfil$\crcr#2\crcr\sim\crcr}}}

\renewcommand{\thefootnote}{\fnsymbol{footnote}}

\begin{document} \begin{titlepage} 
\rightline{\vbox{\halign{&#\hfil\cr
&SLAC-PUB-9192\cr
&April 2002\cr}}}
\begin{center} 
\openup4.0\jot
 
{\Large\bf Kaluza-Klein Effects on Higgs Physics in Universal Extra 
Dimensions}
\footnote{Work supported by the Department of 
Energy, Contract DE-AC03-76SF00515}
\medskip

\normalsize 
{\bf \large Frank J. Petriello}
\vskip .2cm
Stanford Linear Accelerator Center \\
Stanford University \\
Stanford CA 94309, USA\\
\vskip .2cm

\end{center} 
\openup2.0\jot
\begin{abstract} 

We examine the virtual effects of Kaluza-Klein (KK) states on Higgs physics 
in universal extra 
dimension models.  We study the partial widths $\Gamma_{h \rightarrow gg}$, 
$\Gamma_{h \rightarrow \gamma \gamma}$, and $\Gamma_{h \rightarrow \gamma Z}$, 
which are 
relevant for Higgs production and detection in future collider experiments.  
These interactions occur at one loop in the Standard Model, as do 
the KK contributions.  We find that the deviations 
induced by the KK exchanges can be significant; for one extra 
dimension, the $gg \rightarrow h$ production rate is increased by 
$10\% - 85\%$ for the mass of the first KK state in the range $500 \gsim m_1 
\gsim 1500$ 
GeV, a region untested by current direct search and precision 
measurement constraints.  The $h \rightarrow \gamma \gamma$ decay 
width is decreased by $\lsim 20\%$ in the same mass range.  For two or more 
universal extra dimensions the results are cutoff dependent, and can only be 
qualitatively estimated.  We comment on the 
detectability of these shifts at the LHC and at future $e^+ e^-$ and 
$\gamma \gamma$ colliders.

\end{abstract}

\renewcommand{\thefootnote}{\arabic{footnote}} \end{titlepage}


\section{Introduction} 

Theories in which gravity propagates in large extra spacetime 
dimensions~\cite{ADD} have received a great deal of attention during the past 
several years.  These models permit some of 
the qualitative features of string theory, such as the existence of extra 
dimensions and stringy resonances~\cite{Stringy}, to be tested 
experimentally, and predict 
the appearance of a wide variety of phenomenology at future high 
energy colliders~\cite{ADDpheno}.  They also furnish a solution to the 
hierarchy problem 
by lowering the fundamental Planck scale to a TeV.  The string theoretic 
motivation for extra dimensions also allows new dimensions in which Standard 
Model (SM) or other non-gravitational fields can 
propagate~\cite{Antoniadis:1990ew}; for consistency 
with experimental constraints they must have a size of order an 
${\rm TeV}^{-1}$.  Models utilizing this idea have been shown to yield a 
host of interesting phenomena, including TeV-scale unification~\cite{Unif}, 
explanations of fermion Yukawa hierarchies~\cite{Yukawa}, mechanisms for 
generating neutrino masses~\cite{Neutrino}, and methods of rendering axions 
invisible~\cite{Dienes:1999gw}.

One proposed scenario, referred to as the Universal Extra Dimensions 
(UED) model~\cite{Appelquist:2000nn}, allows all the SM fields to propagate in 
${\rm TeV}^{-1}$ extra dimensions.  At tree level, the momentum in the 
extra dimensions is conserved, requiring pair production of the associated 
Kaluza-Klein (KK) modes at colliders and preventing tree level mixing effects 
from altering precision electroweak measurements.  The compactification scale 
of the UED can 
therefore be as low as 300 GeV for one extra dimension, and remains 
less than 1 TeV for two UED.  The phenomenological implications 
of UED for collider experiments~\cite{UEDcol}, $b \rightarrow s 
\gamma$~\cite{Agashe:2001xt}, and the muon anomalous magnetic 
moment~\cite{Appelquist:2001jz} have been studied, and new mechanisms for 
generating neutrino masses~\cite{Appelquist:2002ft} and suppressing proton 
decay~\cite{Appelquist:2001mj} have been developed.

The detection of direct production of UED KK states at future colliders is 
expected to be 
difficult, for the following two reasons: (i) a remnant of extra-dimensional 
momentum conservation 
when loop effects are included implies the existence of a neutral, stable 
KK mode, leading to the necessity of 
interpreting missing energy signatures; (ii) the near degeneracy of the KK 
excitations within each level 
renders the mass shifts due to radiative corrections important in 
determining the pattern of decays~\cite{Chengtalk}.  It is therefore 
interesting to determine 
whether there are other, indirect ways in which the effects of UED can be 
detected.  One such possibility is through the modification of Higgs 
production and decay processes at future colliders; determining whether such 
deviations can significantly modify Higgs properties is also important 
considering 
the necessity of establishing the mechanism of electroweak symmetry-breaking.  
We study here the processes $gg \rightarrow h$, $h \rightarrow \gamma \gamma$, 
and $h \rightarrow \gamma Z$; the first interaction is the dominant Higgs 
production mechanism at the LHC, while the second is the primary discovery 
mode for $m_H \lsim 150$ GeV.  All three processes occur at one loop in 
the SM, the same order at which the KK excitations first contribute; we 
expect, and find, 
that these effects are quite large for the low compactification scales 
allowed for 
UED.  Furthermore, graviton exchanges do not contribute to these processes at 
one loop, as can be seen from the unitary gauge Feynman rules 
in~\cite{ADDpheno}.  
We can therefore with some sense of security neglect the gravitational 
effects which presumably also appear in the complete theory in which the UED 
are embedded~\cite{Hall:1999fe}.  We concentrate here on modifications 
arising from physics in UED, rather than from other extra-dimensional models, 
for two reasons: (i) in the Randall-Sundrum 
model~\cite{Randall:1999ee} in which SM fields can propagate in the full 
5-dimensional spacetime, 
Higgs physics is already modified at tree-level by the mixing between the 
Higgs and the radion field which stabilizes the extra 
dimension~\cite{Hewett:2002nk}; (ii) in 
${\rm TeV}^{-1}$ models where only gauge fields propagate in the bulk, we 
expect the effects to be unobservable, because the bound on the 
compactification scale from electroweak precision fits is quite 
high~\cite{Masip:1999mk,Rizzo:1999br} and the 
top quark KK excitations which induce the majority of the effects found 
here are absent.  

Our paper is organized as follows.  In Sec. 2 we review the formulation of 
the SM in one additional UED, focusing on the appropriate gauge-fixing and 
the mixing within the top quark KK tower.  We study the modifications of the 
processes $gg \rightarrow h$, $h \rightarrow \gamma \gamma$, and 
$h \rightarrow \gamma Z$ in Sec. 3; we find that the heavy KK modes decouple, 
yielding finite, unambiguous results for one UED.  For more than one 
UED the sums over KK modes diverge, and only qualitative statements can be 
made.  We find that observable 
modifications to Higgs production and decay processes occur for 
compactification masses $m_1 \lsim 1.5$ TeV; the $gg \rightarrow h$ 
production rate is increased by $\approx 10\%-85\%$ for $1500 \gsim m_1 
\gsim 500$ GeV, respectively, while the decay widths are shifted by 
$\lsim 20\%$ in the same interval.  We present our conclusions in Sec. 4.

\section{Kaluza-Klein Reduction of the 5-dimensional Standard Model}

We review here the formulation of the UED model, 
in which all the SM fields can propagate in the extra 
dimensions.  We restrict our attention to the 5-dimensional scenario, and 
focus on the issues most pertinent to our calculation: the appropriate choice 
of gauge-fixing and the effects of mixing within the top quark KK tower.  A 
detailed construction of the SM in UED is given 
in~\cite{Appelquist:2000nn}, while a discussion of generalized $R_{\xi}$ 
gauges in a variety of extra-dimensional models is presented 
in~\cite{Muck:2001yv}.

We begin with the action
\begin{eqnarray}
S &=& \int_{-\pi R}^{\pi R} dy \int d^{4}x \, \bigg\{ \, -\frac{1}{2} 
\sum_{i=1}^{3}  
{\rm Tr} \left[  F_{i MN}F_{i}^{MN} \right] + (D_{M}H)^{\dagger} D^{M}H  
+ \mu^2 |H|^2 -\frac{\lambda_5}{4} |H|^4  \nonumber \\ & &
+ i \, \bar{Q} \!\not\!{D} Q + i \, \bar{t} \!\not\!{D} \, t 
+ \left[ \lambda^{t}_{5} \bar{Q} \, i\, \sigma_2 \, H^{*} \, t +{\rm h.c.} 
\right] \bigg\} \,\, .
\label{KKaction}
\end{eqnarray}
Here $(M,N)$ are the 5-dimensional Lorentz indices, and $R$ is the radius of 
the fifth dimension, which we have anticipated compactifying on $S^1 /Z_2$.  
$H$ is the Higgs doublet, 
and the $F_{i}^{MN}$ are the field strengths for the SM gauge groups.  $Q$ is 
the third generation quark doublet and $t$ is the top quark 
singlet; we will not need the remaining SM fermions in our analysis, and they 
have consequently not been included.  The covariant derivative $D_M$ can be 
expressed as
\begin{equation}
D_{M} = \partial_{M} - i\sum_{i=1}^{3} g^{i}_{5}\, T^{a}_{i} A^{a}_{i\,M} 
\,\, ,
\end{equation}
where the $g^{i}_{5}$ are the 5-dimensional coupling constants for $U(1)_Y$, 
$SU(2)_{L}$, and $SU(3)_c$, and the $T^{a}_{i}$ are the generators of these 
groups.  The 5-dimensional Dirac matrices are $\gamma^{M} = (\gamma^{\mu},
i\, \gamma^{5})$.  $\mu^2$, $\lambda_5$, and $\lambda^{t}_5$ are the 
5-dimensional versions of the usual Higgs couplings and top quark Yukawa 
coupling.  The parameters $\lambda_5$, $\lambda^{t}_5$, and $g^{i}_{5}$ are 
dimensionful, and must be rescaled to obtain the correct dimensionless SM 
couplings; no rescaling is necessary for the the Higgs mass parameter $\mu^2$.

To derive the 4-dimensional effective action we must expand the 5-dimensional 
fields into their KK modes; we must also remove several extra massless  
particles from the resulting theory.  Five-dimensional fermions are 
necessarily vector-like, and we wish to obtain the chiral zero modes 
necessary for construction of the SM; this necessitates the removal of the 
extra zero modes appearing in the top quark KK tower.  We also must eliminate 
the zero modes of the scalars $A_{i}^{5}$ that arise in the reduction of the 
gauge fields.  To do this we follow the standard recipe of compactifying the 
fifth dimension on an $S^{1} /Z_{2}$ orbifold and requiring that the fields 
whose zero modes we wish to remove are odd under the orbifold projection 
$y \rightarrow -y$.  The appropriate KK expansions of the 5-dimensional fields 
are:
\begin{eqnarray}
H(x^{\mu},y) &=& \frac{1}{\sqrt{2\pi R}} \left\{ H^{(0)}(x^{\mu}) +\sqrt{2} 
\sum_{n=1}^{\infty} H^{(n)}(x^{\mu}) \, {\rm cos}\, (\frac{ny}{R}) \right\} 
\,\, , \nonumber \\
A_{i\, \mu}(x^{\nu},y) &=& \frac{1}{\sqrt{2\pi R}} \left\{ A^{(0)}_{i\, \mu}
(x^{\nu}) +\sqrt{2} \sum_{n=1}^{\infty} A^{(n)}_{i\, \mu}(x^{\nu}) \, 
{\rm cos}\, (\frac{ny}{R}) \right\} \,\, , \nonumber \\
A_{i}^{5}(x^{\nu},y) &=& \frac{1}{\sqrt{\pi R}} \sum_{n=1}^{\infty} 
A^{5(n)}_{i} (x^{\nu}) \, {\rm sin}\, (\frac{ny}{R}) \,\, , \nonumber \\
Q(x^{\nu},y) &=&  \frac{1}{\sqrt{2\pi R}} \bigg\{ Q^{(0)}_{L}(x^{\nu}) + 
\sqrt{2} \sum_{n=1}^{\infty} \bigg[ P_{L} Q^{(n)}_{L}(x^{\nu}) \, {\rm cos}\, 
(\frac{ny}{R}) \nonumber \\ & & 
+ P_{R} Q^{(n)}_{R}(x^{\nu}) \, {\rm sin}\, (\frac{ny}{R}) 
\bigg] \bigg\} \,\, , \nonumber \\
t(x^{\nu},y) &=&  \frac{1}{\sqrt{2\pi R}} \bigg\{ t^{(0)}_{R}(x^{\nu}) + 
\sqrt{2} \sum_{n=1}^{\infty} \bigg[ P_{R} t^{(n)}_{R}(x^{\nu}) \, {\rm cos}\, 
(\frac{ny}{R}) \nonumber \\ & & 
+ P_{L} t^{(n)}_{L}(x^{\nu}) \, {\rm sin}\, (\frac{ny}{R}) 
\bigg] \bigg\} \,\, ,
\label{KKexp}
\end{eqnarray}
where we have introduced the projection operators $P_{R,L}=(1 \pm 
\gamma^{5})/2$.  We thus obtain the desired zero modes $A^{(0)}_{i\, \mu}$, 
$Q^{(0)}_{L}$, and $t^{(0)}_{R}$, corresponding to the SM fields.  These 
expansions should be inserted into 
the action of Eq.~\ref{KKaction}.  We must also expand the zero mode Higgs 
doublet around its vev, and express the KK Higgs doublets in terms of their 
component fields:
\begin{equation}
H^{(0)}=\left( \begin{array}{c} \phi^{(0)+} \\ \frac{1}{\sqrt{2}} 
\left( \nu +h^{(0)} +i\, \chi^{(0)} \right) \end{array} \right) \,\, , 
\,\,\,\, 
H^{(n)}=\left( \begin{array}{c} \phi^{(n)+} \\ \frac{1}{\sqrt{2}} 
\left( h^{(n)} +i\, \chi^{(n)} \right) \end{array} \right) \,\, .
\label{higgsexp}
\end{equation}
Here $\nu$ is the usual 4-dimensional Higgs vev, $h^{(0)}$ is the physical 
zero mode Higgs, and $\chi^{(0)}$, $\phi^{\pm (0)}$ are the zero mode 
Goldstone bosons.  The $h^{(n)}$ are the CP-even Higgs KK excitations, the 
$\chi^{(n)}$ are CP-odd scalars that will combine with the $Z^{5(n)}$ to 
form a tower of CP-odd Higgs bosons and generate the longitudinal components 
for the $Z^{(n)}_{\mu}$, and the $\phi^{\pm (n)}$ are charged scalars that 
together with the $W^{\pm 5(n)}$ will form a tower of charged Higgs scalars 
and longitudinal components for the $W^{\pm (n)}_{\mu}$.  Inserting the 
expansions of Eqs.~\ref{KKexp} and~\ref{higgsexp} into the action in 
Eq.~\ref{KKaction} leads to a slew of mass terms, mixings, and couplings.  
We focus first on the masses and mixings in the gauge sector, introducing 
the appropriate gauge-fixing terms and deriving the spectrum of physical 
states and Goldstone fields.

We first examine the photon KK tower; the relevant mass terms and mixings are
\begin{equation}
S^{A} = \int d^{4}x \sum_{n=1}^{\infty} \left\{ \frac{1}{2}m_{n}^{2} 
A^{(n)}_{\mu}A^{\mu(n)} -m_{n}A^{(n)}_{\mu}\partial^{\mu}A^{5(n)} \right\}
\,\, ,
\label{Amix}
\end{equation}
where $m_{n} = n/R$ is the KK mass of the $n$th level arising from the 
derivative $\partial_5$ acting on the 5-dimensional wavefunctions of 
Eq.~\ref{KKexp}.  The most natural choice of gauge-fixing is the 
five-dimensional analog of the Feynman gauge,
\begin{equation}
S^{A}_{gf}=-\frac{1}{2} \int_{-\pi R}^{\pi R} dy \int d^{4}x \, 
\left(\partial_{M} A^{M} \right)^2 \,\, .
\label{Agf}
\end{equation}
Utilizing the KK expansion of $A^{M}$, and summing Eqs.~\ref{Amix} 
and~\ref{Agf}, we find that the mixing between $A^{(n)}_{\mu}$ and 
$A^{5(n)}$ cancels, and that we are left with the mass terms
\begin{equation}
S^{A}+S^{A}_{gf} = \frac{1}{2}\int d^{4}x  \, \sum_{n=1}^{\infty} \left\{ 
m_{n}^2 A^{(n)}_{\mu}A^{\mu(n)} -m_{n}^2 \left(A^{5(n)} \right)^2 \right\} 
\,\, ;
\end{equation}
the spectrum then consists of a massless zero mode $A^{(0)}_{\mu}$, a tower 
of KK 
modes $A^{(n)}_{\mu}$ with masses $m_{n}$, and a tower of Goldstone particles 
$A^{5(n)}$ also with mass $m_{n}$.  The treatment of the gluon KK tower 
proceeds identically, and we will not present it explicitly.

We next study the $Z$ boson KK tower, together with the KK excitations of the 
zero mode Goldstone particle, $\chi$.  The corresponding masses and mixing 
terms are
\begin{eqnarray}
S^{Z}&=&\frac{1}{2} \int d^{4}x \bigg\{ M_{Z}^2 \left(Z^{(0)} \right)^2 
+2M_{Z}Z^{(0)}_{\mu}\partial^{\mu}\chi^{(0)} +\sum_{n=1}^{\infty} \bigg[
-m_{n}^2 \left(\chi^{(n)}\right)^2 +m_{Z,n}^2 Z^{(n)}_{\mu}Z^{\mu(n)}
\nonumber \\ & & -M_{Z}^2 \left( Z^{5(n)} \right)^2 -2m_{n}M_{Z}Z^{5(n)}
\chi^{(n)} +2Z^{(n)}_{\mu}\partial^{\mu}\left( M_{Z}\chi^{(n)} 
-m_{n}Z^{5(n)} \right) \bigg] \bigg\} \,\, ,
\label{Zmix}
\end{eqnarray}
where we have introduced the abbreviation $m_{Z,n}^2 = M_{Z}^2 +m_{n}^2$.
We choose the straightforward 5-dimensional generalization of the usual SM 
Feynman gauge,
\begin{equation}
S^{Z}_{gf}= -\frac{1}{2}\int_{-\pi R}^{\pi R} dy \int d^{4}x \, 
\left(\partial_{M} Z^{M}-M_{Z}\chi \right)^2 \,\, ;
\end{equation}
utilizing the KK expansion of Eq.~\ref{KKexp} and combining this with 
Eq.~\ref{Zmix}, we derive the following mass terms:
\begin{eqnarray}
S^{Z}+S^{Z}_{gf}&=&\frac{1}{2} \int d^{4}x \, \bigg\{ M_{Z}^2 Z^{(0)}_{\mu} 
Z^{\mu (0)} - M_{Z}^2 \left( \chi^{(0)} \right)^2 
+\sum_{n=1}^{\infty} \bigg[ m_{Z,n}^2 Z^{(n)}_{\mu}Z^{\mu(n)} \nonumber \\
& & - m_{Z,n}^2 \left( \chi^{(n)} \right)^2 -m_{Z,n}^2 \left(Z^{5(n)}\right)^2
\bigg] \bigg\} \,\, ;
\end{eqnarray}
the mixing between $Z^{(n)}_{\mu}$ and $Z^{5(n)}$ cancels.  It is clear from 
Eq.~\ref{Zmix} that the linear combinations of 
fields that serve as Goldstone modes for the $Z$ boson KK tower are 
\begin{equation}
G^{(n)}_{Z} =\frac{M_{Z}\chi^{(n)}-m_{n}Z^{5(n)}}{\sqrt{M_{Z}^2 +m_{n}^2}}
\,\, ,
\end{equation}
while the physical CP-odd scalars are
\begin{equation}
\chi^{(n)}_{Z} = \frac{m_{n}\chi^{(n)}+M_{Z}Z^{5(n)}}{\sqrt{M_{Z}^2 +m_{n}^2}}
\,\, .
\end{equation}
With the gauge choice we have made, the states $G^{(n)}_{Z}$, 
$\chi^{(n)}_{Z}$, and $Z^{(n)}_{\mu}$ all possess the mass $m_{Z,n}$.

Finally, we consider the masses and mixing terms involving the $W^{\pm}$ KK 
tower and the KK excitations of the zero mode Goldstone fields $\phi^{\pm}$:
\begin{eqnarray}
S^{W}&=& \int d^{4}x \, \bigg\{ M_{W}^2 W^{+(0)}_{\mu}W^{- \mu (0)} 
+ iM_{W} \left( W^{-(0)}_{\mu}\partial^{\mu}\phi^{+(0)} -
W^{+(0)}_{\mu}\partial^{\mu}\phi^{-(0)} \right) \nonumber \\ & & 
+\sum_{n=1}^{\infty} \bigg[ -m_{n}^2 \phi^{+(n)}\phi^{-(n)} 
+m_{W,n}^2 W^{+(n)}_{\mu}W^{- \mu (n)} -M_{W}^2 W^{+5(n)}W^{-5(n)} \nonumber \\
& &  -im_{n}M_{W} \left( W^{-5(n)} 
\phi^{+(n)} -W^{+5(n)} \phi^{-(n)} \right) 
-W^{-(n)}_{\mu} \partial^{\mu} \left( m_n W^{+5(n)} -iM_{W}\phi^{+(n)} \right)
\nonumber \\ & &
-W^{+(n)}_{\mu} \partial^{\mu} \left( m_n W^{-5(n)} +iM_{W}\phi^{-(n)} \right)
\bigg] \bigg\} \,\, ,
\label{Wmix}
\end{eqnarray}
where we have abbreviated $m_{W,n}^2 = M_{W}^2 +m_{n}^2$.  The 
appropriate choice of gauge-fixing term is again the obvious 5-dimensional 
extension of the SM Feynman gauge:
\begin{equation}
S^{W}_{gf}= - \int_{-\pi R}^{\pi R} dy \int d^{4}x \, 
\left( \partial_{M}W^{+M} -iM_{W} \phi^{+} \right) \left( \partial_{M}W^{-M} 
+iM_{W} \phi^{-} \right) \,\, .
\end{equation}
Inserting the KK expansions of Eq.~\ref{KKexp} into this expression, and 
adding it to Eq.~\ref{Wmix}, we find that the mixing between 
$W^{\pm (n)}_{\mu}$ and $W^{\pm 5(n)}$ cancels, and we obtain the mass terms  
\begin{eqnarray}
S^{W}+S^{W}_{gf}&=& \int d^{4}x \bigg\{ M_{W}^2 W^{+(0)}_{\mu}W^{-\mu (0)} 
-M_{W}^2 \phi^{+(0)}\phi^{-(0)} \nonumber \\ & & \!\!
+\sum_{n=1}^{\infty} \bigg[ m_{W,n}^2 W^{+(n)}_{\mu}W^{-\mu (n)} 
-m_{W,n}^2 W^{+5(n)}W^{-5(n)} -m_{W,n}^2 \phi^{+(n)}\phi^{-(n)} 
\bigg] \bigg\} \,\, .
\end{eqnarray}
Again, the Goldstone modes are linear combinations of the 5-dimensional 
components of the gauge fields, $W^{\pm 5(n)}$, and the KK excitations of 
the zero mode Goldstone, $\phi^{\pm (n)}$:
\begin{equation}
G^{\pm (n)} = \frac{m_{n}W^{\pm 5(n)} \mp iM_{W}\phi^{\pm (n)}}{\sqrt{
m_{n}^2 + M_{W}^2}} \,\, .
\end{equation}
The physical charged Higgs pair is the orthogonal combination:
\begin{equation}
H^{\pm (n)} = \frac{m_{n}\phi^{\pm (n)} \mp iM_{W}W^{\pm 5(n)}}{\sqrt{
m_{n}^2 + M_{W}^2}} \,\, .
\label{Hcharge}
\end{equation}
In the 5-dimensional generalization of the SM Feynman gauge we employ,
the fields $W^{\pm \mu (n)}$, $G^{\pm (n)}$, and $H^{\pm (n)}$ share the 
common mass $m_{W,n}$.

Having computed the spectrum of states in the gauge sector, we can now derive 
the interactions of the gauge and scalar particles; we identify the 
4-dimensional couplings as $\lambda = \lambda_5 /2\pi R$, $\lambda^{t} = 
\lambda^{t}_{5} / \sqrt{2\pi R}$, and $g^{i} = g^{i}_{5} / \sqrt{2\pi R}$ so 
that the zero mode interactions match those of the SM.  Letting 
$\phi^{(n)}_{i}$ denote a KK excitation of an 
arbitrary SM field, the contributing KK interactions take the form 
\begin{equation}
\phi^{(0)}_{i}\phi^{(n)}_{j}\phi^{(n)}_{k} \,\, , \,\,\,\,
\phi^{(0)}_{i}\phi^{(0)}_{j}\phi^{(n)}_{k}\phi^{(n)}_{l} \,\, .
\end{equation}
The explicit expressions for these vertices are simple to obtain; for every SM 
vertex $\phi^{(0)}_{i}\phi^{(0)}_{j}\phi^{(0)}_{k}$ or 
$\phi^{(0)}_{i}\phi^{(0)}_{j}\phi^{(0)}_{k}\phi^{(0)}_{l}$, there is a 
corresponding KK vertex with exactly the same coupling strength.  We note 
explicitly that the $h^{(0)}W^{+(n)}W^{-(n)}$ and $h^{(0)}Z^{(n)}Z^{(n)}$ 
vertices are identical to the $h^{(0)}W^{+(0)}W^{-(0)}$ and 
$h^{(0)}Z^{(0)}Z^{(0)}$ vertices; the masses that appear in the KK 
interactions are $M_{W}$ and $M_{Z}$, not $m_{W,n}$ and $m_{Z,n}$.  The heavy 
KK states decouple from the processes considered here, allowing us to obtain 
finite results in five dimensions when the sum over KK modes is performed.  
The only other interactions needed for our calculation are those involving 
$W^{\pm 5(n)}$; these are given in the appendix.  For a complete list of the 
SM vertices we refer the reader to~\cite{Denner:kt}; we will not reproduce 
them here.  

We now derive the interactions of the top quark KK states required in our 
analysis.  Although there is no mixing between different levels of the top 
quark KK tower, the doublet and singlet states within each level mix.  The 
mass matrix for the $n$th KK level arising from the reduction of 
Eq.~\ref{KKaction} is 
\begin{equation}
\left( \bar{Q}^{(n)}_{L} \, , \,\, \bar{t}^{\, (n)}_{L} \right)
\left( \begin{array}{cc} m_{n} & m_{t} \\ m_{t} & -m_{n} \end{array} \right)
\left( \begin{array}{c} Q^{(n)}_{R} \\ t^{(n)}_{R} \end{array} \right) 
+ {\rm h.c.} \,\, ,
\end{equation}
where $m_{t}$ is the zero mode top quark mass.  This can be diagonalized with 
the following unitary matrices for the left and right-handed fields:
\begin{equation}
U_{L}^{(n)} = \left( \begin{array}{cc} {\rm cos}(\alpha^{(n)} /2) & {\rm sin}(
\alpha^{(n)} /2)
\\ {\rm sin}(\alpha^{(n)} /2) & -{\rm cos}(\alpha^{(n)} /2) \end{array} 
\right) \,\, , 
\,\,\,\, 
U_{R}^{(n)} = \left( \begin{array}{cc} {\rm cos}(\alpha^{(n)} /2) & {\rm sin}(
\alpha^{(n)} /2)
\\ -{\rm sin}(\alpha_{(n)} /2) & {\rm cos}(\alpha^{(n)} /2) \end{array} 
\right) \,\, ,
\end{equation}
where both states in the physical basis have mass 
$m_{t,n}=\sqrt{m_{n}^2 + m_{t}^2}$, and ${\rm cos}(\alpha^{(n)})=m_{n}/
m_{t,n}$, ${\rm sin}(\alpha^{(n)})=m_{t}/m_{t,n}$.  We must derive the 
couplings of these states to $h^{(0)}$, $A^{(0)}_{\mu}$, $Z^{(0)}_{\mu}$, 
and $g^{(0)}_{\mu}$ for our analysis.  
Denoting the mass eigenbasis of the $n$th level by the vector $T^{(n)}$, we 
find the following KK interactions:
\begin{eqnarray}
S^{t} &=& \int d^{4}x \sum_{n=1}^{\infty} \bigg\{ \bar{T}^{(n)} 
\!\not\!\!{A}^{(0)} 
C^{(n)}_{A} T^{(n)} + \bar{T}^{(n)} \!\not\!{Z}^{(0)} C^{(n)}_{Z,V} T^{(n)}
+ \bar{T}^{(n)} \!\not\!{g}^{(0)} C^{(n)}_{g} T^{(n)} \nonumber \\ & &
+ \left[ h^{(0)} \bar{T}^{(n)} C^{(n)}_{h} T^{(n)} +{\rm h.c.} \right] \bigg\}
\,\, .
\end{eqnarray}
The coupling matrices appearing in this expression are 
\begin{eqnarray}
C^{(n)}_{A}&=&  e Q_{t} \, \left( 
\begin{array}{cc} 
1 & 0 \\ 0 & 1 \end{array} \right) \, , \,\,\,\, 
C^{(n)}_{Z,V} = \frac{g}{c_{W}} \, \left( \begin{array}{cc} g_{v}-g_{a} \,
{\rm cos}(\alpha^{(n)}) & 0\\ 0 
& g_{v}+g_{a} \, {\rm cos}(\alpha^{(n)}) \end{array} \right) \nonumber \\
C^{(n)}_{h} &=& m_{t} \, \left( \begin{array}{cc} {\rm sin}(\alpha^{(n)}) & 
{\rm cos}(\alpha^{(n)}) \\ -{\rm cos}(\alpha^{(n)}) & {\rm sin}(\alpha^{(n)})
 \end{array} \right) \, , \,\,\,\,
C^{(n)}_{g}=  g_{3} \, \left( 
\begin{array}{cc} 
1 & 0 \\ 0 & 1 \end{array} \right) \,\, ,
\label{Tcoup}
\end{eqnarray}
where $Q_{t}$ is the top quark charge in units of $e$, $c_{W}$ is the cosine 
of the weak mixing angle, $g$ and $g_3$ are respectively the coupling 
constants of $SU(2)_L$ and $SU(3)_c$, and $g_{v}$, $g_{a}$ are the top quark 
vector and axial couplings to the SM $Z$.  The $\gamma^{5}$ component of 
the $Z$ interaction does not contribute to the studied processes, and has 
consequently not been included.  We note that the coupling of top quark KK 
states to the Higgs is proportional to $m_{t}$, not $m_{t,n}$; the heavy KK 
top quarks decouple as do the $W^{\pm (n)}$ and $Z^{(n)}$ towers.  This is in 
contrast to the behavior of a heavy fourth generation quark, whose coupling 
to the Higgs is proportional to its mass, and which does not decouple.

We now possess the tools required to study corrections to Higgs boson 
production and decay processes arising from one loop KK exchanges.  We will 
concentrate on the processes $gg \rightarrow h$, $h \rightarrow \gamma 
\gamma$, and $h \rightarrow \gamma Z$, which occur at one loop in the SM; the 
KK contributions to these interactions are therefore of the same order as the 
SM contributions.  The decoupling of the higher KK modes allows us to obtain 
finite predictions 
when only one extra dimension is considered; furthermore, at one loop graviton 
exchanges do not contribute to these processes, rendering our 
neglect of the gravity sector of the theory justifiable.  These features 
allow us to obtain unambiguous and testable predictions.

\section{KK Effects in One Loop Higgs Processes}

We now study the effects of virtual KK exchanges in $gg \rightarrow h$, 
$h \rightarrow 
\gamma \gamma$, and $h \rightarrow \gamma Z$, processes relevant for Higgs 
production and decay at the LHC.  Both the SM and KK contributions to these 
interactions occur at one loop, and we therefore expect the modifications
arising from KK exchanges to be significant.  This is indeed the case; we will 
find that KK effects are visible for the compactification mass $m_1$ in the 
range 
$ 400 \, {\rm GeV} \lsim m_1 \lsim 1500 \, {\rm GeV}$, a region consistent 
with the constraints arising from both direct searches and precision 
measurements~\cite{Appelquist:2000nn}.

\subsection{ $ gg \rightarrow h$}

The process $gg \rightarrow h$ proceeds in the SM through diagrams containing 
fermion triangle loops.  We consider only contributions arising from the top 
quark and its KK tower; the couplings of other fermions to the Higgs are 
much smaller than that of the top quark, and are negligible in our analysis.  
The production cross section, which is proportional to the $h \rightarrow gg$ 
width, can be written in the form 
\begin{equation}
\sigma_{gg \rightarrow h} = \frac{G_F \left[ \alpha_{s}(m_H) \right]^2}{32 
\sqrt{2} \pi m_{H}^4}| F_{t} |^2 \,\, ,
\end{equation}
where $G_F$ is the Fermi constant, $\alpha_s(m_{H})$ is the QCD coupling 
strength evaluated at the Higgs mass scale, and $F_t$ is the contribution 
of the loop integrals over the top quark KK tower contributions.  Introducing 
the abbreviation 
\begin{equation}
C_{0} (m^2) = C_{0} (m_{H}^2,0,0;m^2,m^2,m^2)
\label{scalar}
\end{equation}
for the three-point scalar Passarino-Veltman 
function~\cite{Passarino:1978jh,Denner:kt}, the SM result becomes 
\begin{equation}
F_{t}^{SM} = -2m_{t}^2 + m_{t}^2 \left( m_{H}^2 - 4m_{t}^2 \right) 
C_{0} (m_{t}^2) \,\, .
\label{FtSM}
\end{equation}
The scalar three-point function of Eq.~\ref{scalar} can be evaluated in terms 
of elementary functions, yielding
\begin{equation}
C_{0}(m^2) = \left\{ \begin{array}{l}
-\frac{2}{m_{H}^2} \left[ {\rm arcsin} \left( \frac{1}{\sqrt{\tau}} \right) 
\right]^2 \,\,\,\,\,\,\,\,\,\,\,  \tau \geq 1 \\ 
\\
\frac{1}{2m_{H}^2} \left[ {\rm ln}\left( \frac{1+\sqrt{1-\tau}}
{1-\sqrt{1-\tau}} \right) -i\pi \right]^2 
\,\,\,\, \tau < 1 \end{array} \right. \,\, ,
\end{equation}
where $\tau = 4m^2 / m_{H}^2$.  The couplings of the top quark KK excitations 
to both zero mode Higgs bosons 
and photons are given in Eq.~\ref{Tcoup}; utilizing these expressions, we can 
write $F_{t}=F_{t}^{SM}+F_{t}^{KK}$, where
\begin{equation}
F_{t}^{KK} = 2 \, m_{t} \sum_{n=1}^{\infty} \, m_{t,n} \, 
{\rm sin}(\alpha^{(n)})
\left\{ -2+ \left( m_{H}^2 - 4m_{t,n}^2 \right)
C_{0} (m_{t,n}^2)  \right\} \,\, .
\label{FtKK}
\end{equation}
In obtaining this formula, and other formulae presented in this paper, we 
used QGRAF~\cite{Nogueira:1991ex} to check that 
we included all the appropriate diagrams and FORM~\cite{Vermaseren:2000nd} to 
verify our algebraic manipulations.  In the limit $m^2 \rightarrow \infty$, 
$C_{0}(m^2) \approx - 1/2m^2 - m_{H}^2 / 24m^4$.  
Applying this result to $F_{t}^{KK}$, we find that
\begin{equation}
F_{t}^{KK} \approx - \frac{2 m_{H}^2 m_{t}^2}{3} \sum_{n=1}^{\infty} 
\frac{1}{m_{n}^2} \,\, 
\end{equation}
in the limit that the KK mass parameters $m_n$ are much larger than either 
$m_{t}$ or $m_{H}$.  In five dimensions this sum is over a single index $n$, 
and we obtain a convergent result.  In greater than five dimensions we must 
sum over an array of indices $n_{i}$, where $i$ ranges over the number of 
extra dimensions, and $F_{t}^{KK}$ diverges (in more than five dimensions 
there is also a greater multiplicity of states arising from the KK 
reduction~\cite{Cheng:1999fu}, which affects both the finite piece and the 
coefficient of the divergent part of the sum).  These divergent sums can be 
evaluated by 
introducing a cutoff $\Lambda$; in six dimensions, for example, this leads to 
the result $F_{t}^{KK} \propto {\rm ln}(\Lambda R)$.  We will not study 
scenarios with $D > 5$ here; we expect, however, that the results we obtain in 
the 5-dimensional case will be qualitatively similar to those found in the 
complete $D>5$ theory in which these UED are embedded, and in which this 
arbitrariness is removed.

\begin{figure}[htbp]
\centerline{
\psfig{figure=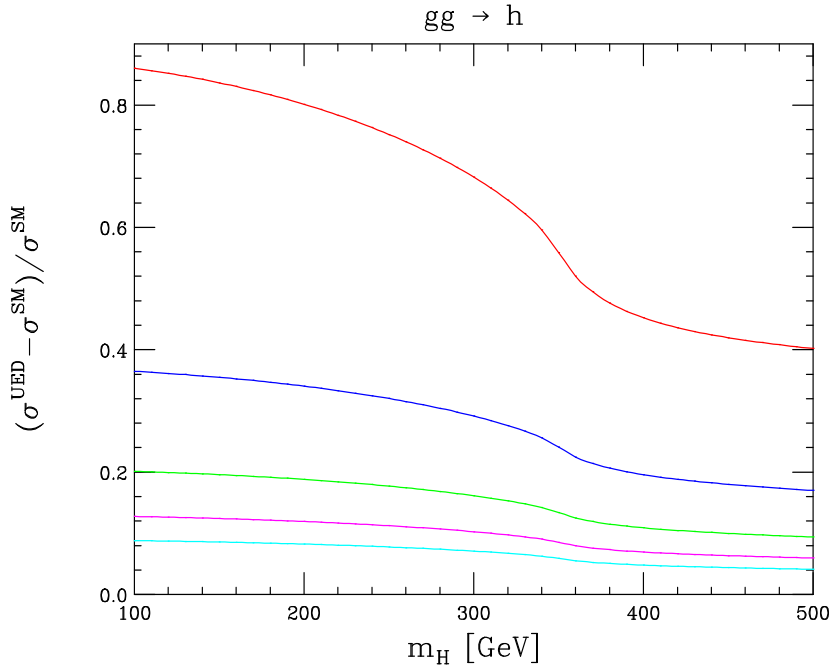,height=8.8cm,width=10.9cm,angle=0}}
\caption{The fractional deviation of the $gg \rightarrow h$ production rate 
in the UED model as a function of $m_H$; from top to bottom, the results are 
for $m_1 = 500,750,1000,1250,1500$ GeV.}
\label{ggh}
\end{figure}

Since the lower bound on the KK mass parameter $m_1$ in UED models is quite 
low, $m_1 \gsim 300-400$ GeV, we expect the deviations due to the virtual KK 
exchanges to be large.  We present in 
Fig.~\ref{ggh} the fractional deviation of the production cross section from 
that of the SM for the following choices 
of compactification mass: $m_1 = 500,750,1000,1250,1500$ GeV.  It was argued 
in~\cite{Appelquist:2000nn} that the 4-dimensional effective theory remains 
valid until $m_n \approx 10$ TeV.  A negligible fraction of the effects 
found here are induced by KK modes with masses above this value, and we can 
therefore trust our results for the compactification masses $m_1 \leq 1.5$ TeV 
considered.  Fits to the 
electroweak precision data within the framework of extra-dimensional models 
typically allow Higgs masses larger than the 95\% CL upper bound obtained in 
the SM~\cite{Rizzo:1999br}; we therefore present results for the range 
$m_H \leq 500$ GeV.  For $m_1 = 500$ GeV and $m_H \approx 120$ GeV, the 
production rate is $ \approx 85\%$ larger than in the SM; this decreases to 
$40\%$ for a 500 GeV Higgs.  For $m_1 = 1500$ GeV and $m_H \approx 120$ GeV 
the increase is $\approx 10\%$.  A more complete analysis would take into 
account the next-to-leading order QCD corrections, which significantly 
increase $\sigma_{gg \rightarrow h}$~\cite{Gghnlo}, 
and the next-to-next-to-leading order QCD corrections in the $m_t \rightarrow 
\infty$ limit, which have recently been computed~\cite{Gghnnlo}.  However, the 
KK contributions will receive the same QCD corrections, and we expect that 
for $m_1 \lsim 1.5$ TeV and a light Higgs boson, deviations arising from 
physics in UED should be observable.  Future 
$e^+ e^- $ linear colliders will determine the $h \rightarrow gg$ 
decay width with a $10\%-12.5\% $ precision for Higgs masses in the range 
$120-140$ GeV~\cite{Abe:2001wn}, indicating that 
compactification masses $m_1 \lsim 1500$ GeV are indeed testable.

We will examine 
the observability of UED contributions at future colliders in more detail in 
the following subsections, where we compute the corresponding deviations 
arising from KK 
exchanges in the decays $h \rightarrow \gamma \gamma$ and $h \rightarrow 
\gamma Z$.  This will allow us to estimate the total shift in production 
rates for the processes $gg \rightarrow \gamma \gamma, \gamma Z$, which 
are relevant for Higgs searches at the LHC.

\noindent

\subsection{ $h \rightarrow \gamma \gamma$}

We now study the decay $h \rightarrow \gamma \gamma$, which is the primary 
discovery mode at the LHC for a Higgs with mass $m_H \lsim 150$ GeV.  At one 
loop, this process proceeds through both top quark and gauge sector 
loops, with the latter involving the $W^{\pm}$ tower and its associated 
Goldstone modes and ghosts.  The decay width can be written as 
\begin{equation}
\Gamma_{h \rightarrow \gamma \gamma} = \frac{G_{F} \alpha^2}{8 \sqrt{2} 
\pi^3 m_H } |F|^2 \,\, ,
\end{equation}
where $\alpha$ is the electromagnetic coupling, and $F = F_{W}+3 Q_{t}^2 
F_t$.  The SM result for $F_{t}^{SM}$ is given in Eq.~\ref{FtSM}, and 
\begin{equation}
F_{W}^{SM} = \frac{1}{2}m_{H}^2 +3 M_{W}^2 -3M_{W}^2 \left( m_{H}^2 - 
2M_{W}^2 \right)  C_{0}(M_{W}^2) \,\, .
\end{equation}
In the UED model there are additional contributions from the top quark KK 
tower, the $W^{\pm}$ tower and its associated Goldstone modes, ghost KK 
states, and the $H^{\pm}$ tower defined in Eq.~\ref{Hcharge}.  We set 
$F_t = F_{t}^{SM}+F_{t}^{KK}$ and $ F_{W}=F_{W}^{SM}+F_{G}^{KK}$, with 
$F_{t}^{KK}$ denoting the top quark KK tower contribution and $F_{G}^{KK}$ 
including the contributions of the KK excitations in the gauge and Higgs 
sectors. $F_{t}^{KK}$ is then given by the expression in Eq.~\ref{FtKK}, and 
\begin{equation}
F_{G}^{KK} = \sum_{n=1}^{\infty} \left\{ \frac{1}{2} m_{H}^2 +4 M_{W}^2 
-\left[ 4M_{W}^2 \left( m_{H}^2 -2 m_{W,n}^2 \right) -m_{H}^2 m_{W,n}^2 
\right] C_{0}(m_{W,n}^2) \right\} \,\, .
\end{equation}
Using the expansion $C_{0}(m^2) \approx 
- 1/2m^2 - m_{H}^2 /24m^4$, it is simple to check that this sum converges in 
five dimensions.  
However, it diverges for $D > 5$, as does the $gg \rightarrow h$ production 
cross section.  We again expect that the results we obtain will be 
qualitatively similar for $D>5$ when the cutoff dependence is fixed by a more 
complete theory.

\begin{figure}[htbp]
\centerline{
\psfig{figure=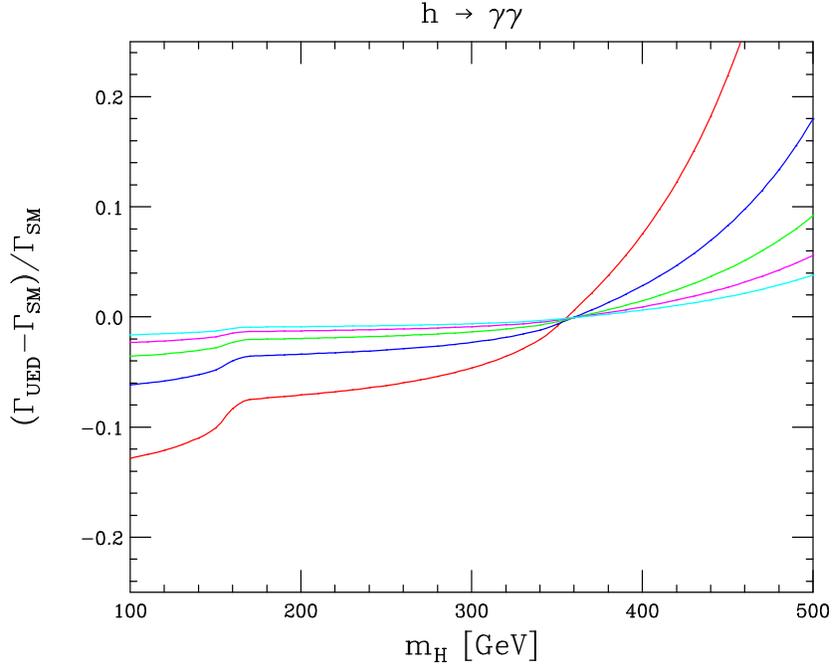,height=8.8cm,width=10.9cm,angle=0}}
\caption{The fractional deviation of the $h \rightarrow \gamma \gamma$ decay 
width in the UED model as a function of $m_H$; from top to bottom on the 
right, the results are for $m_1 = 500,750,1000,1250,1500$ GeV.}
\label{hgg}
\end{figure}

\begin{figure}[htbp]
\centerline{
\psfig{figure=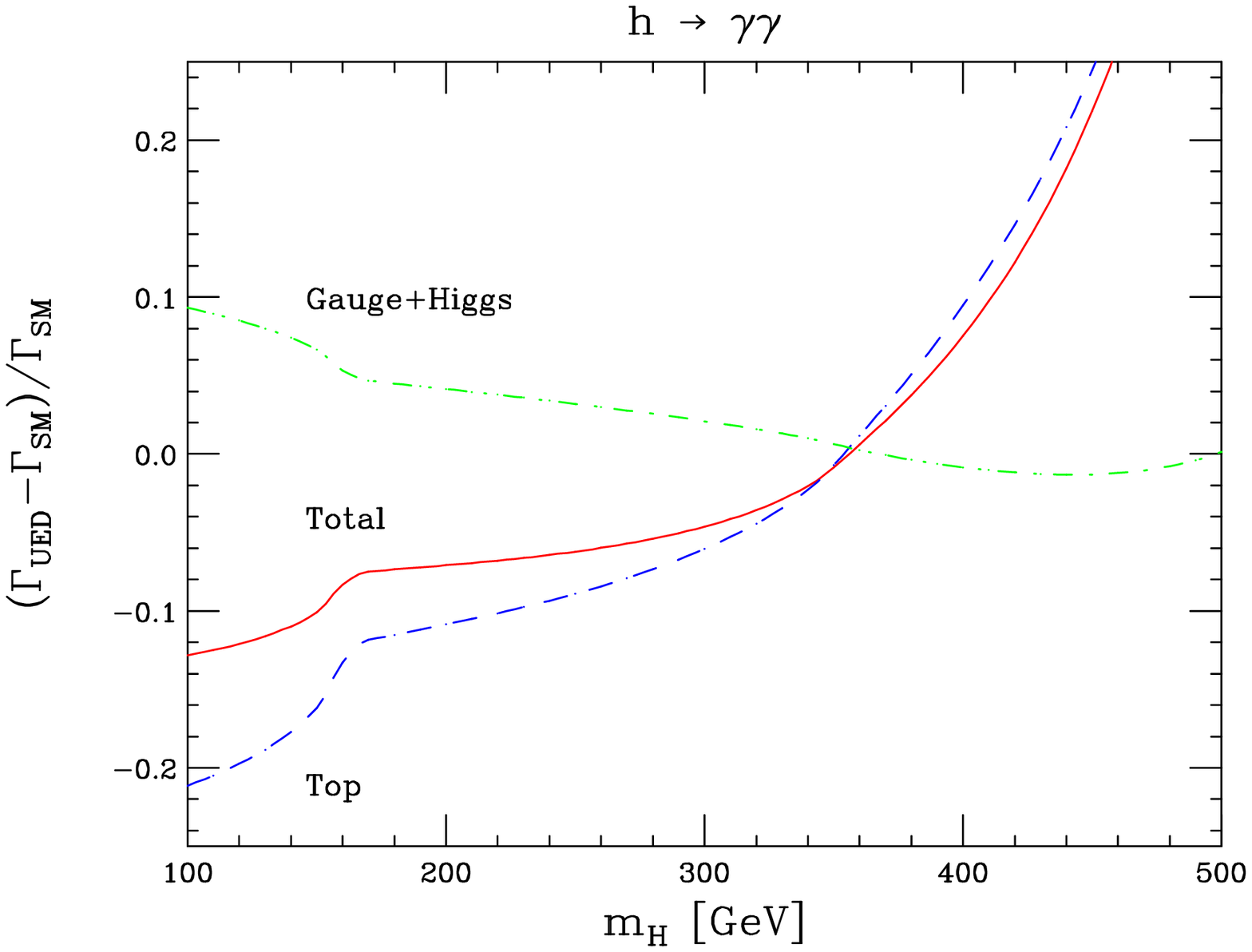,height=8.8cm,width=10.9cm,angle=0}}
\caption{The fractional deviation of the $h \rightarrow \gamma \gamma$ 
decay width
for $m_1 = 500$ GeV as a function of $m_H$, with the contributions of the top 
quark sector and the gauge and Higgs sectors shown separately.}
\label{hggcomp}
\end{figure}

The interference between the SM and KK contributions is more intricate in 
$h \rightarrow \gamma \gamma$ than in $gg \rightarrow h$, as thresholds exist 
at both $2M_W$ and $2m_t$ where the relative importance of the various 
contributions can change.  The fractional 
deviation of the $h \rightarrow \gamma \gamma$ decay width is shown in 
Fig.~\ref{hgg} for five choices of $m_1$, and the fractional deviations due to 
the top quark KK tower and the gauge and Higgs tower contributions are 
presented 
separately for $m_1 = 500$ GeV in Fig.~\ref{hggcomp}.  The $\gamma \gamma$ 
decay width in the 
UED model is $\approx 12\%$ smaller than in the SM for $m_H \lsim 2 M_W$ 
and $m_1 = 500$ GeV, the Higgs mass region in which this decay is expected to 
be the discovery mode at the LHC; this result drops to $\approx 4\%$ for 
$m_1 = 1000$ GeV.  However, at $m_H \approx 2m_t$ the decay width in the UED 
scenario becomes larger than in the SM.  The relevant contributions of the 
top quark and gauge sector KK towers are shown in Fig.~\ref{hggcomp}.  The 
contribution of the top quark KK tower, the dominant UED term, interferes 
destructively with the SM result below the $2m_t$ threshold; this behavior 
reverses above threshold.

\noindent

\noindent

\begin{figure}[htbp]
\centerline{
\psfig{figure=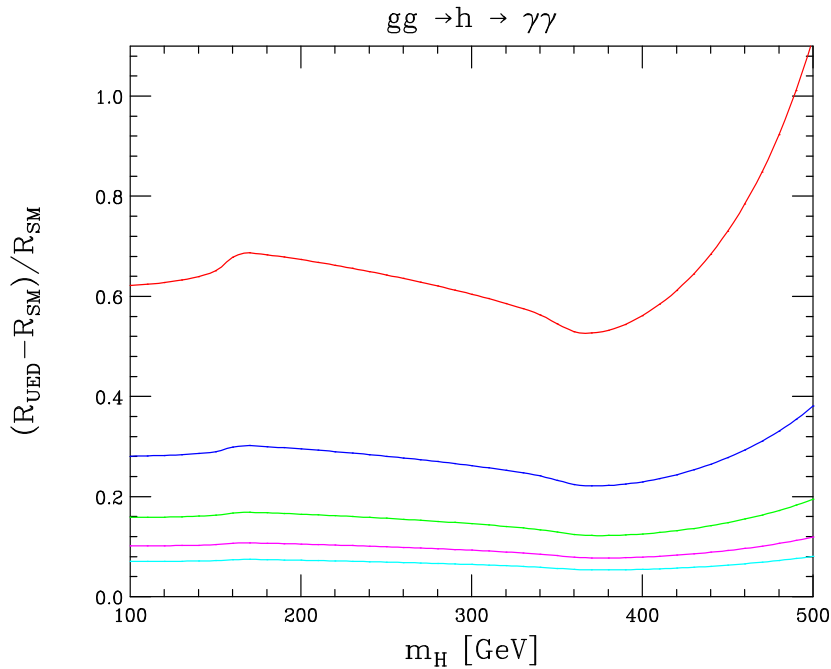,height=8.8cm,width=10.9cm,angle=0}}
\caption{The fractional deviation of $R=\sigma_{gg \rightarrow h} \times 
\Gamma_{h \rightarrow 
\gamma \gamma}$, the $\gamma \gamma$ production rate,
in the UED model as a function of $m_H$; from top to bottom, the 
results are for $m_1 = 500,750,1000,1250,1500$ GeV.}
\label{gammarate}
\end{figure}

\vspace*{-0.5in}
To determine the sensitivity of the LHC to these effects, we must compute 
the net shift in the $\gamma \gamma$ production rate resulting from the 
deviations
in both $gg \rightarrow h$ and $h \rightarrow \gamma \gamma$.   For resonant 
production of the Higgs, the $\gamma \gamma$ signal is well approximated 
by taking $\sigma_{gg \rightarrow h} \times \Gamma_{h \rightarrow 
\gamma \gamma}$, including the parton density functions evaluated at the 
relevant scale, and multiplying by the appropriate prefactors.  The fractional 
deviation in the $\gamma \gamma$ production rate is presented in 
Fig.~\ref{gammarate} for five choices of $m_1$.  For $m_H \lsim 150 $ GeV, 
the region of interest at the LHC, the increase in $\sigma_{gg \rightarrow h}$ 
and the decrease in $\Gamma_{h \rightarrow \gamma \gamma}$ yield a 
total $\approx 10\%-65\%$ increase in the total rate as $m_1$ is varied from 
1250 GeV to 500 GeV, respectively.  The LHC is expected to be sensitive to 
this rate at 
the $10\%-15\%$ level~\cite{Zeppenfeld:2002ng}; consequently, we expect 
signals from UED to be visible if $m_1 \lsim 1000-1250$ GeV.  An independent 
measurement of the $h \rightarrow \gamma \gamma$ decay width will be 
achievable 
at future linear colliders; for $m_h \lsim 150$ GeV, a measurement of the 
$h\gamma\gamma$ coupling at the $7\% -10\% $ level will be 
possible~\cite{Abe:2001wn}.  This will provide a test of UED models with 
$m_1 \lsim 800$ GeV.  A measurement of the $h \rightarrow \gamma \gamma $ 
width with an accuracy of $\approx 2\%$ is possible with the proposed photon 
collider option of future $e^+ e^-$ colliders~\cite{Badelek:2001xb}; this 
would allow probes of the UED model with KK mass parameter $m_1 \lsim 
1500$ GeV.

\noindent

\subsection{ $h \rightarrow \gamma Z$}

We examine here the decay $h \rightarrow \gamma Z$, which proceeds in the SM 
through top quark and gauge sector loops.  Although the width of this process 
exceeds the $h \rightarrow \gamma \gamma$ width for Higgs masses in the 
range $m_H \gsim 130$ GeV, the single photon and need to demand a leptonic 
$Z$ decay for reconstruction purposes render it less interesting at the LHC.  
However, since it potentially provides another test of the detailed 
properties of the Higgs boson, we study modifications of this decay arising 
from physics in UED.

The decay width can be expressed as 
\begin{equation}
\Gamma_{h \rightarrow \gamma Z} = \frac{\alpha G_{F}^2 M_{W}^2 m_{H}^3 s_{W}^2
}{64 \pi^4} \left(1-\frac{M_{Z}^2}{m_{H}^2} \right)^3 |F|^2 \,\, ,  
\end{equation}
where $s_{W}$ is the sine of the weak mixing angle.  We introduce the 
abbreviation
\begin{eqnarray}
C_{2}(m^2) &=& C_{1}(m_{H}^2,M_{Z}^2,0;m^2,m^2,m^2) +
C_{11}(m_{H}^2,M_{Z}^2,0;m^2,m^2,m^2) \nonumber \\ & & 
+ C_{12}\, (m_{H}^2,M_{Z}^2,0;m^2,m^2,m^2)
\,\, ,
\end{eqnarray}
where $C_{1}$, $C_{11}$, and $C_{12}$ are the Passarino-Veltman tensor 
coefficients defined in~\cite{Denner:kt}, and change slightly our shorthand 
notation for the scalar three-point function:
\begin{equation}
C_{0}(m^2) = C_{0}(m_{H}^2,M_{Z}^2,0;m^2,m^2,m^2) \,\, .
\end{equation}
These can be evaluated in terms of elementary functions~\cite{Djouadi:1996yq};
setting $\tau_Z = 4m^2 /M_{Z}^2$ and $\tau_H = 4m^2 /m_{H}^2$, we have
\begin{eqnarray}
4m^2 C_{2}(m^2) &=& \frac{\tau_Z \tau_H}{2 \left( \tau_Z -\tau_H \right)} 
+ \frac{\tau_Z \tau_{H}^2}{2 \left( \tau_Z -\tau_H \right)^2} 
\left\{ \tau_Z \left[ f(\tau_Z) - f(\tau_H) \right] +2 \left[ g(\tau_Z) - 
g(\tau_H) \right] \right\} \,\, , \nonumber \\ 
4m^2 C_{0}(m^2) &=& \frac{2 \tau_Z \tau_H}{\tau_Z -\tau_H} \left[ f(\tau_Z) 
- f(\tau_H) \right] \,\, ,
\end{eqnarray}
where 
\begin{equation}
f(\tau) = \left\{ \begin{array}{l}
\left[ {\rm arcsin} \left( \frac{1}{\sqrt{\tau}} \right) 
\right]^2 \,\,\,\,\,\,\,\,\,\,\,  \tau \geq 1 \\ 
\\
-\frac{1}{4} \left[ {\rm ln}\left( \frac{1+\sqrt{1-\tau}}
{1-\sqrt{1-\tau}} \right) -i\pi \right]^2 
\,\,\,\, \tau < 1 \end{array} \right. \,\, ,
\end{equation}
and
\begin{equation}
g(\tau) = \left\{ \begin{array}{l}
 \sqrt{\tau -1} \,{\rm arcsin} \left( \frac{1}{\sqrt{\tau}} \right) 
 \,\,\,\,\,\,\,\,\,\,\,  \tau \geq 1 \\ 
\\
\frac{1}{2} \sqrt{1-\tau} \left[ {\rm ln}\left( \frac{1+\sqrt{1-\tau}}
{1-\sqrt{1-\tau}} \right) -i\pi \right] 
\,\,\,\, \tau < 1 \end{array} \right. \,\, .
\end{equation}
Writing 
\begin{equation}
F = {\rm cot}(\theta_W ) F_{W} +3 \frac{g_v Q_{t}}{s_W c_W} F_{t}  \,\, ,
\end{equation}
where $\theta_W$ is the weak mixing angle, the SM result becomes
\begin{eqnarray}
F_{t}^{SM} &=& 4 \, m_{t}^2 \left\{ 4 C_{2}(m_{t}^2) +C_{0}(m_{t}^2) 
\right\} 
\nonumber \\ 
F_{W}^{SM} &=& 4 \, \bigg\{ -M_{W}^2 \left(3 C_{0}(M_{W}^2) + 
5C_{2}(M_{W}^2) 
\right) -\frac{1}{2}m_{H}^2 C_{0}(M_{W}^2) \nonumber \\ & &
+ s_{W}^2 \left[ M_{W}^2 \left( 
6 C_{2}(M_{W}^2) +4 C_{0}(M_{W}^2) \right) +m_{H}^2 
C_{2}(M_{W}^2) \right] \bigg\} \,\, .
\end{eqnarray}
In our notation, $g_v = I_{3} /2 - s_{W}^2 Q_{t} $, where $I_3$ is the third 
component of the top quark weak 
isospin. In the UED model, there are additional contributions from both 
the top quark KK tower and the gauge sector KK excitations; partitioning these 
pieces as in the $h \rightarrow \gamma \gamma$ case, $F_{t}=F_{t}^{SM} + 
F_{t}^{KK}$ and $F_{W}=F_{W}^{SM}+F_{G}^{KK}$, we find 
\begin{eqnarray}
F_{t}^{KK} &=& 8 \, \sum_{n=1}^{\infty} \left\{ m_{t} \, m_{t,n} \, 
{\rm sin}(\alpha^{(n)}) \left[ 4 C_{2}(m_{t,n}^2) +C_{0}(m_{t,n}^2) 
\right] \right\} \nonumber \\ 
F_{G}^{KK} &=& 4 \, \sum_{n=1}^{\infty} \bigg\{-M_{W}^2 \left(
3 C_{0}(M_{W}^2) + 7C_{2}(M_{W}^2) 
\right) -\frac{1}{2}m_{H}^2 C_{0}(M_{W}^2) \nonumber \\ & &
+ s_{W}^2 \left[ M_{W}^2 \left( 
8 C_{2}(M_{W}^2) +4 C_{0}(M_{W}^2) \right) +m_{H}^2 
C_{2}(M_{W}^2) \right] \bigg\} \,\, .
\end{eqnarray}
We have used the couplings given in Eq.~\ref{Tcoup} in deriving these results.
It can be checked that 
these sums converge in $D=5$, but diverge for $D>5$; again, we concentrate on 
the $D=5$ scenario.

\begin{figure}[htbp]
\centerline{
\psfig{figure=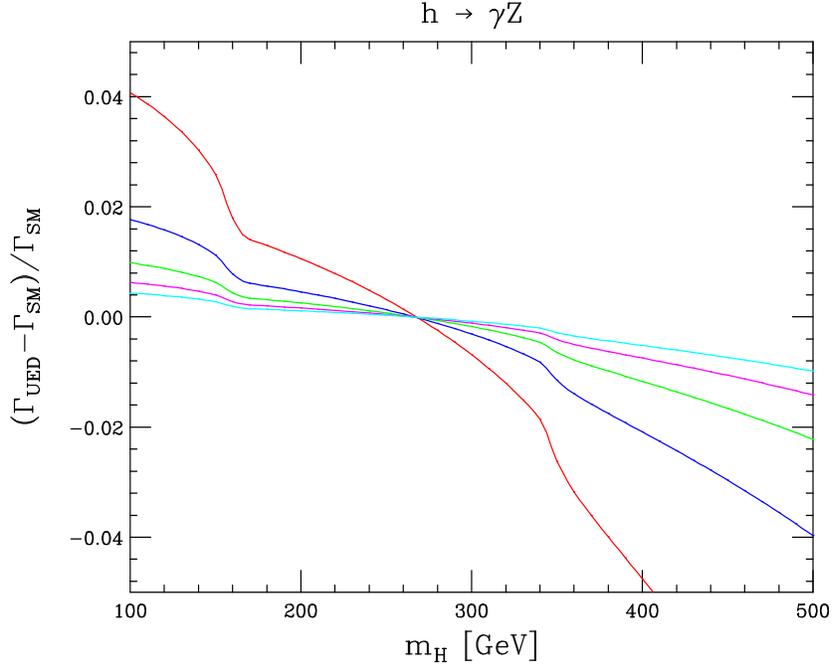,height=8.8cm,width=10.9cm,angle=0}}
\caption{The fractional deviation of the $h \rightarrow \gamma Z$ decay 
width in the UED model as a function of $m_H$; from top to bottom on the 
left, the results are for $m_1 = 500,750,1000,1250,1500$ GeV.}
\label{hgz}
\end{figure}

\begin{figure}[htbp]
\centerline{
\psfig{figure=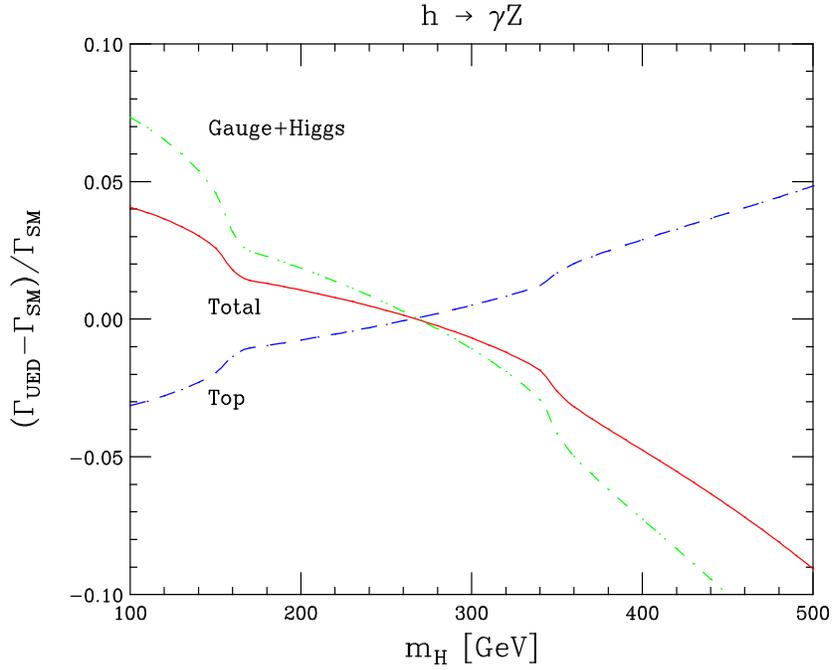,height=8.8cm,width=10.9cm,angle=0}}
\caption{The fractional deviation of the $h \rightarrow \gamma Z$ 
decay width
for $m_1 = 500$ GeV as a function of $m_H$, with the contributions of the top 
quark sector and the gauge and Higgs sectors shown separately.}
\label{hgzcomp}
\end{figure}

We present the fractional deviation of $\Gamma_{h \rightarrow \gamma Z}$ in 
the UED model in Fig.~\ref{hgz}  for five choices of the KK mass parameter 
$m_1$, and show the relative contributions of the top quark and gauge sectors 
in Fig.~\ref{hgzcomp}.  The decay width in the UED model is slightly larger 
than the SM width for $m_H \lsim 275$ GeV, and slightly smaller for higher 
values of $m_H$.  The top quark and gauge sector KK towers have contributions 
with approximately equal magnitude but opposite sign, as seen in 
Fig.~\ref{hgzcomp}, and their effects tend to cancel.  For all $m_H$ and 
$m_1$ considered the deviation is $\lsim 10\%$, and is hence smaller than the 
modifications to the $gg$ and 
$\gamma \gamma$ widths.  An effect of this magnitude is possibly observable 
at future linear colliders, although a detailed analysis of this 
decay mode has not been performed; it is also possible that such an 
effect could be observed in the $\gamma \gamma$ collision option of 
future colliders.

\noindent

\noindent

\begin{figure}[htbp]
\centerline{
\psfig{figure=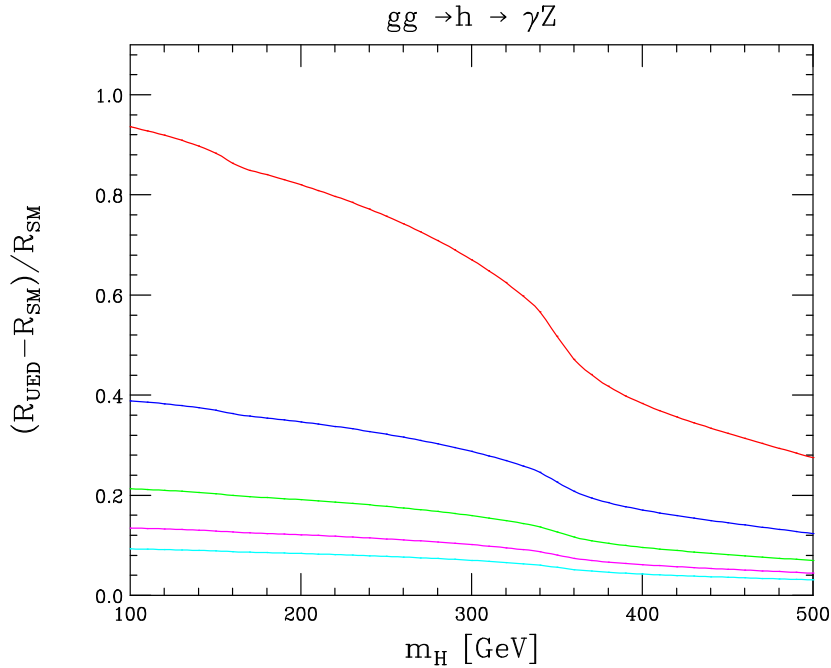,height=8.8cm,width=10.9cm,angle=0}}
\caption{The fractional deviation of $R=\sigma_{gg \rightarrow h} \times 
\Gamma_{h \rightarrow 
\gamma Z}$, the $\gamma Z$ production rate,
in the UED model as a function of $m_H$; from top to bottom, the 
results are for $m_1 = 500,750,1000,1250,1500$ GeV.}
\label{gzrate}
\end{figure}

\vspace*{-1.5cm}
The fractional deviation of the $\gamma Z$ production rate at the LHC via 
$gg \rightarrow h \rightarrow \gamma Z$ is shown in Fig.~\ref{gzrate} for 
five choices of $m_H$.  The production increase is $\approx 95\%$ for 
$m_H \lsim 150$ GeV and $m_1 = 500$ GeV, and $\approx 20\%$ for 
$m_1 = 1000$ GeV.  This shift is caused primarily by the $gg 
\rightarrow h$ deviation; however, it may render this decay mode visible 
above the background at the LHC.  Again, a detailed analysis of this decay 
at the LHC has not been performed.

\noindent

\section{Conclusions}

We have studied the virtual effects of KK excitations in UED on Higgs 
production and decay 
processes relevant for high energy experiments at the LHC and at future 
linear colliders.  The heavy KK modes decouple, allowing us to obtain 
unambiguous predictions for one extra dimension.  For two or more extra 
dimensions the KK 
mode sums diverge, and while we expect our results in these scenarios to be 
qualitatively similar to those obtained here, we cannot make precise 
predictions.  We have found that 
the KK excitation contributions can be quite significant; the $gg \rightarrow 
h$ production rate can be $\approx 85\%$ larger than the SM result for 
$m_H \lsim 150$ GeV and KK mass 
parameter $m_1 = 500$ GeV, a value allowed by current constraints.  For 
$m_1 \approx 1500$ GeV, the rate increase is $\approx 10\%$; assuming the SM 
theoretical prediction is under control by the time the LHC turns on, this 
should be an observable shift.  The corresponding deviations in the 
$h \rightarrow gg$ decay width can be probed at future $e^+ e^-$ colliders, 
allowing compactification masses in the range $m_1 \lsim 1500$ GeV to be 
tested.  The width of the decay $h \rightarrow \gamma 
\gamma$, the 
primary discovery mode for a Higgs with mass $m_H \lsim 150$ GeV at the LHC, 
is decreased relative to the SM prediction by $\approx 12\%$.  The total 
$\gamma \gamma$ production rate is increased by 
$\approx 10\%-65\%$ for $1250 \gsim m_1 \gsim 500$ GeV when the 
reaction $gg \rightarrow h \rightarrow \gamma \gamma$ relevant at the LHC is 
considered.  With the 
$10\%-15\%$ accuracy expected in the determination of this rate at the LHC, 
compactification masses $m_1 \lsim 1250$ GeV can be probed.  The 
$h \rightarrow 
\gamma \gamma$ width can be independently measured at future $e^+ e^-$ and 
$\gamma \gamma$ colliders; we expect the effects from compactification masses 
$m_1 \lsim 800$ 
GeV to be observable with the $7\%-10\%$ precision expected at $e^+ e^-$ 
colliders, and from masses $m_1 \lsim 1500$ GeV to be testable with the 
$2\%$ precision expected at $\gamma \gamma$ colliders.  Finally, we have 
examined 
the deviations in the decay $h \rightarrow \gamma Z$ predicted by UED models.  
We have found that the 
deviations from the SM result are less than $\approx 10\%$ throughout the 
$m_H,m_1$ region studied.  However, the process $gg \rightarrow h 
\rightarrow \gamma Z$ is expected to increase by $\approx 20\% - 95\% $ 
for compactification masses in the range $1000 \gsim m_1 \gsim 500$ GeV.  No 
detailed study of this process has been performed for either the LHC or 
future linear colliders; however, the production increase at the LHC is 
possibly visible above background.

How do these results compare with the deviations induced in other new 
physics models?  In the Randall-Sundrum model studied in~\cite{Hewett:2002nk}, 
where the Higgs and radion fields mix, both $\Gamma_{h \rightarrow gg}$ and 
$\Gamma_{h \rightarrow \gamma\gamma}$ are decreased throughout the allowed 
parameter space; the total Higgs production rate can be decreased to 
$\lsim 1\%$ of the SM value for a large range of Higgs-radion mixing 
strengths.  These results for $\Gamma_{h \rightarrow gg}$ are the opposite of 
those found here, in which the width is increased throughout the allowed 
parameter space.  The situation is murkier in the Minimal 
Supersymmetric Standard Model (MSSM), as a large number of parameters enter 
calculations at the one loop level.  A detailed study of Higgs physics in 
the MSSM was performed in~\cite{Carena:2001bg}.  Typically, $\Gamma_{h 
\rightarrow gg}$ is decreased by $\lsim 15\%$ throughout the allowed 
parameter space, while $\Gamma_{h \rightarrow \gamma\gamma}$ is shifted by 
$\lsim 5\%$, with the direction of the shift parameter dependent.  Again, the 
deviation in $\Gamma_{h \rightarrow gg}$ is opposite that found here.  
$\Gamma_{h \rightarrow gg}$ was also studied 
in~\cite{Cacciapaglia:2001nz} within a supersymmetric 
extra-dimensional scenario~\cite{Barbieri:2000vh}.  In this model, 
$\Gamma_{h \rightarrow gg}$ 
receives contributions from loops of both top and stop KK excitations; the 
localization of fermion Yukawa couplings at orbifold fixed points induces 
mixing within these KK towers.  The width is decreased relative to its 
SM value throughout the entire parameter space; for a Higgs with $m_H \approx 
120$ GeV the width is $ \lsim 25\%$ of the SM result.  This is again opposite 
the shift found here.  The effects of any of these scenarios on Higgs physics 
should therefore be 
distinguishable from the shifts found in the UED model studied here; the 
direct 
production of the various new states associated with each model should also 
assist in distinguishing them.

In summary, the virtual effects of KK excitations in UED can significantly 
alter Higgs properties which will be measured  at 
future colliders.  The implications of radiative corrections in 
extra-dimensional models have not been studied extensively, primarily 
because of the resulting divergences.  We have 
shown that 
in certain scenarios such effects are both calculable and important, and we 
believe that further investigations along these lines should be 
undertaken.

\noindent
{\Large\bf Acknowledgements}

\noindent
It is a pleasure to thank C. Anastasiou, J. Hewett, K. Melnikov, and T. Rizzo 
for many helpful suggestions and conversations.  We also thank 
G. Cacciapaglia, M. Cirelli, and G. Cristadoro for informing us of their 
results in~\cite{Cacciapaglia:2001nz}.  This work was supported in 
part by the National Science Foundation Graduate Research Program.

\newpage
\noindent
{\Large\bf Appendix}

\noindent
We present here the Feynman rules involving the $W^{\pm 5(n)}$,  which are 
relevant for the 
calculations in this paper.  All momenta are assumed to flow into the vertices.
 
\vspace*{-2.5cm}
\begin{figure}[b]
\begin{picture}(350,180)(10,-30)

\Line(10,48.28)(50,48.28)
\put(10,28){$h^{(0)}$}
\Line(50,48.28)(78.28,76.56)
\put(82,68.28){$W^{\pm 5(n)}$}
\Line(50,48.28)(78.28,20)
\put(80,18){$\phi^{\mp (n)}$}
\put(150,48.28){$= \mp \, \frac{e}{2s_{W}} \, m_{n}$}

\Line(10,146.56)(50,146.56)
\put(10,156.28){$h^{(0)}$}
\Line(50,146.56)(78.28,174.84)
\put(82,166.56){$W^{+ 5(n)}$}
\Line(50,146.56)(78.28,118.28)
\put(80,116.28){$W^{- 5(n)}$}
\put(150,146.56){$= -i\frac{e}{s_{W}} M_{W}$}

\Photon(10,244.84)(50,244.84){5}{3}
\put(-5,264.56){$ \left\{ A^{(0)}_{\mu},Z^{(0)}_{\mu} \right\} $}
\Photon(50,244.84)(78.28,273.12){5}{3}
\put(82,264.84){$W^{\pm (n)}_{\nu}$}
\Line(50,244.84)(78.28,216.56)
\put(80,214.56){$W^{\mp 5(n)}$}
\put(150,244.84){ $= \mp \, m_{n} \left\{ 1, -\frac{c_{W}}{s_{W}} \right\} 
g_{\mu \nu} $}

\Photon(10,343.12)(50,343.12){5}{3}
\put(-5,362.84){ $ \left\{ A^{(0)}_{\mu},Z^{(0)}_{\mu} \right\} $ }
\put(10,329.84){$p_3$}
\Line(50,343.12)(78.28,371.4)
\put(69,348.12){$p_1$}
\put(82,363.12){$W^{+ 5(n)}$}
\Line(50,343.12)(78.28,314.84)
\put(80,312.84){$W^{- 5(n)}$}
\put(69,332.84){$p_2$}
\put(150,343.12){ $= -ie \left\{ 1, -\frac{c_{W}}{s_{W}}
\right\} \left(p_1 - p_2 \right)_{\mu} $}

\Line(51.72,473.12)(80,444.84)
\put(10,464.84){$W^{\pm 5(n)}$}
\Line(51.72,416.56)(80,444.84)
\put(10,414.56){$W^{\mp 5(n)}$}
\Photon(80,444.84)(108.28,473.12){5}{3}
\put(112,464.84){$A^{(0)}_{\mu}$}
\Photon(80,444.84)(108.28,416.56){5}{3}
\put(110,414.56){$ \left\{ A^{(0)}_{\nu},Z^{(0)}_{\nu} \right\} $}
\put(150,444.84){ $= 2ie^2 \left\{ 1, -\frac{c_{W}}{s_{W}} \right\}
 g_{\mu \nu}$}

\end{picture}
\end{figure}

\end{document}